\documentclass[aps,amssymb,amsmath,showpacs,twocolumn,prl,superscriptaddress,preprintnumbers]{revtex4-1}
\usepackage[english]{babel}
\usepackage{amsmath,amsfonts,amssymb,amscd,amsthm}
\usepackage[utf8]{inputenc}
\usepackage{epsfig}
\usepackage{graphicx}
\usepackage{xcolor}

\usepackage{array}
\usepackage[ulem=normalem]{changes}

\usepackage{natbib}

\usepackage{hyperref}
\hypersetup{
    colorlinks=true,                
    breaklinks=true,                
    urlcolor= blue,  
    linkcolor= blue, 
    bookmarksopen=false,
    filecolor=black,
    citecolor=blue, 
    linkbordercolor=blue
}

\begin{document}


\title{Search for a neutron dark decay in \texorpdfstring{\textsuperscript{6}}{6}He}

\author{M.~Le Joubioux}
\email{marius.lejoubioux@ganil.fr}
\affiliation{Grand Acc\'el\'erateur National d'Ions Lourds (GANIL), CEA/DRF-CNRS/IN2P3, Bd.~Henri Becquerel, 14076 Caen, France}

\author{H.~Savajols}
\email{hervé.savajols@ganil.fr}
\affiliation{Grand Acc\'el\'erateur National d'Ions Lourds (GANIL), CEA/DRF-CNRS/IN2P3, Bd.~Henri Becquerel, 14076 Caen, France}

\author{W.~Mittig}
\affiliation{Facility for Rare Isotope Beams, Michigan State University, East Lansing, Michigan 48824, USA}
\affiliation{Department of Physics and Astronomy, Michigan State University, East Lansing, Michigan 48824, USA}

\author{X.~Fl\'echard}
\affiliation{Université de Caen Normandie, ENSICAEN, CNRS/IN2P3, LPC Caen UMR6534, F-14000 Caen, France}

\author{L. Hayen}
\affiliation{Université de Caen Normandie, ENSICAEN, CNRS/IN2P3, LPC Caen UMR6534, F-14000 Caen, France}
\affiliation{Department of Physics, North Carolina State University, Raleigh, 27607 North Carolina, USA}

\author{Yu.E.~Penionzhkevich}
\affiliation{Flerov Laboratory of Nuclear Reactions, Joint Institute for Nuclear Research, Dubna 141980, Russia}
\affiliation{National Research Nuclear University MEPHI, Moscow 115409, Russia}

\author{D.~Ackermann}
\affiliation{Grand Acc\'el\'erateur National d'Ions Lourds (GANIL), CEA/DRF-CNRS/IN2P3, Bd.~Henri Becquerel, 14076 Caen, France}

\author{C.~Borcea}
\affiliation{Horia Hulubei National Institute for Physics and Nuclear Engineering, Reactorului 30, 077125,
Bucharest-M\u{a}gurele, Romania}

\author{L.~Caceres}
\affiliation{Grand Acc\'el\'erateur National d'Ions Lourds (GANIL), CEA/DRF-CNRS/IN2P3, Bd.~Henri Becquerel, 14076 Caen, France}

\author{P.~Delahaye}
\affiliation{Grand Acc\'el\'erateur National d'Ions Lourds (GANIL), CEA/DRF-CNRS/IN2P3, Bd.~Henri Becquerel, 14076 Caen, France}

\author{F.~Didierjean}
\affiliation{Institut Pluridisciplinaire Hubert Curien, 23 Rue du Loess, 67200 Strasbourg, France}


\author{S.~Franchoo}
\affiliation{Universit\'e Paris-Saclay, CNRS/IN2P3, IJCLab, 91405 Orsay, France}

\author{A.~Grillet}
\affiliation{Institut Pluridisciplinaire Hubert Curien, 23 Rue du Loess, 67200 Strasbourg, France}

\author{B.~Jacquot}
\affiliation{Grand Acc\'el\'erateur National d'Ions Lourds (GANIL), CEA/DRF-CNRS/IN2P3, Bd.~Henri Becquerel, 14076 Caen, France}

\author{M.~Lebois}
\affiliation{Universit\'e Paris-Saclay, CNRS/IN2P3, IJCLab, 91405 Orsay, France}

\author{X.~Ledoux}
\affiliation{Grand Acc\'el\'erateur National d'Ions Lourds (GANIL), CEA/DRF-CNRS/IN2P3, Bd.~Henri Becquerel, 14076 Caen, France}

\author{N.~Lecesne}
\affiliation{Grand Acc\'el\'erateur National d'Ions Lourds (GANIL), CEA/DRF-CNRS/IN2P3, Bd.~Henri Becquerel, 14076 Caen, France}

\author{E.~Li\'enard}
\affiliation{Université de Caen Normandie, ENSICAEN, CNRS/IN2P3, LPC Caen UMR6534, F-14000 Caen, France}

\author{S.~Lukyanov}
\affiliation{Flerov Laboratory of Nuclear Reactions, Joint Institute for Nuclear Research, Dubna 141980, Russia}

\author{O.~Naviliat-Cuncic}
\affiliation{Facility for Rare Isotope Beams, Michigan State University, East Lansing, Michigan 48824, USA}
\affiliation{Department of Physics and Astronomy, Michigan State University, East Lansing, Michigan 48824, USA}
\affiliation{Université de Caen Normandie, ENSICAEN, CNRS/IN2P3, LPC Caen UMR6534, F-14000 Caen, France}

\author{J.~Piot}
\affiliation{Grand Acc\'el\'erateur National d'Ions Lourds (GANIL), CEA/DRF-CNRS/IN2P3, Bd.~Henri Becquerel, 14076 Caen, France}

\author{A.~Singh}
\affiliation{Grand Acc\'el\'erateur National d'Ions Lourds (GANIL), CEA/DRF-CNRS/IN2P3, Bd.~Henri Becquerel, 14076 Caen, France}

\author{V.~Smirnov}
\affiliation{Flerov Laboratory of Nuclear Reactions, Joint Institute for Nuclear Research, Dubna 141980, Russia}

\author{C.~Stodel}
\affiliation{Grand Acc\'el\'erateur National d'Ions Lourds (GANIL), CEA/DRF-CNRS/IN2P3, Bd.~Henri Becquerel, 14076 Caen, France}

\author{D.~Testov}
\affiliation{Extreme Light Infrastructure-Nuclear Physics (ELI-NP)/Horia Hulubei National Institute for Physics and Nuclear Engineering (IFIN-HH), Str. Reactorului 30, 077125 Bucharest-M\u{a}gurele, Romania}
\affiliation{Joint Institute for Nuclear Research, Joliot-Curie 6, 141980 Dubna, Moscow region, Russia}

\author{D.~Thisse}
\affiliation{Universit\'e Paris-Saclay, CNRS/IN2P3, IJCLab, 91405 Orsay, France}

\author{J.C.~Thomas}
\affiliation{Grand Acc\'el\'erateur National d'Ions Lourds (GANIL), CEA/DRF-CNRS/IN2P3, Bd.~Henri Becquerel, 14076 Caen, France}

\author{D.~Verney}
\affiliation{Universit\'e Paris-Saclay, CNRS/IN2P3, IJCLab, 91405 Orsay, France}


\begin{abstract}
\color{black}
Neutron dark decays have been suggested as a solution to the discrepancy between bottle and beam experiments, providing a dark matter candidate that can be searched for in halo nuclei. The free neutron in the final state following the decay of $^6$He into $^4$He $+$ $n$ + $\chi$ provides an exceptionally clean detection signature when combined with a high efficiency neutron detector. Using a high-intensity $^6$He$^+$ beam at GANIL, a search for a coincident neutron signal resulted in an upper limit on a dark decay branching ratio of Br$_\chi \leq 4.0\times10^{-10}$ (95\% C.L.). Using the dark neutron decay model proposed originally by Fornal and Grinstein, we translate this into an upper bound on a dark neutron branching ratio of $\mathcal{O}(10^{-5})$, improving over global constraints by one to several orders of magnitude depending on $m_\chi$.

 \color{black}
\end{abstract}

\maketitle

The existence of modern dark matter was suggested nearly one century ago following puzzling observations of galaxy cluster luminosities and, later, galaxy formation and their rotation curves \cite{RevModPhys.90.045002}. Since then, experimental searches have probed a staggeringly large mass scale for such novel constituents of matter, with popular candidates such as Weakly Interacting Massive Particles (WIMPs), sterile neutrinos and axion(-like) particles, backed by significant theoretical efforts to provide viable dark matter candidates \cite{Roszkowski2018, Irastorza2018, Abazajian2012, battaglieri2017cosmic}. Despite eluding direct observation, the $\Lambda$-Cold Dark Matter ($\Lambda$CDM) model used in cosmology is often referred to as the Standard Cosmological Model for its excellent agreement with data \cite{DiValentino2022, Peebles2003, Abbott2022}, in analogy with its particle physics sibling. As such, the existence of dark matter and its interactions are one of the most pressing open questions in particle and astrophysics and inspires searches through a diverse array of probes \cite{Bertone2018a}.

In 2018, Fornal and Grinstein \cite{fornal18, Fornal2020, Fornal2023} described how the long-standing discrepancy between neutron lifetime measurements performed in bottle and beam experiments \cite{Dubbers2021} could be accommodated by a decay into dark final states with total mass between that of the neutron and 937.993 MeV$/c^2$ (due to $^9$Be stability). Immediately following their proposal, searches for $n\to \chi+\gamma$ \cite{tang2018} and $n\to\chi+e^++e^-$ \cite{sun2018, Klopf2019} ruled out this solution in specific scenarios, joined by strong constraints from observed neutron star masses \cite{mckeen2018,baym2018, Zhou2023}, mirror-neutron searches \cite{broussard2022,stereo2022,n2edm2023} and hydrogen stability \cite{McKeen2023}. 

Halo nuclei decays were recognized as excellent model-independent probes \cite{pfutzner18}, as a neutron dark decay within a nucleus with a (two-)neutron separation energy, $S_{(2)n}$, less than $1.592$ MeV is energetically allowed and no coupling to the electromagnetic field is required. As such, halo decay searches are inherently a factor $1/\alpha$ more sensitive than free neutrons searches, with $\alpha$ the fine-structure constant, and probe more generic dark decay scenarios. Recently, $^{11}$Be has attracted attention because of its relatively long half life of 13.76(7) s and its low neutron separation energy of 501.6(3) keV. Several attempts to determine the $^{11}$Be $\rightarrow$ $^{10}$Be rate were performed  \cite{riisager14,riisager20}, completed by measurements of the $\beta-p$ decay channel \cite{ayyad22,lopez22} in order to disentangle the contribution of a hypothetical dark decay branch. Both methods show significant discrepancies, however, underlining the need for other systems to be investigated.

The isotope of choice in this work is $^6$He, a loosely bound unstable nucleus with a two-neutron separation energy of $S_{2n} = 975.45(5)$~keV and lifetime of about 807 ms \cite{Kanafani22}. Removing one neutron to reach $^5$He requires about 1.7 MeV, the unbound ground state of which has a lifetime of about 648~keV \cite{kondev21}. The ground state of $^6$He decays predominantly by $\beta$ decay to the ground state of $^6$Li with a decay energy $Q_{\beta}$ = 3.508~MeV. The only other decay channel energetically allowed is $\alpha + d$ with $Q_{\beta d}$ $=$ 2.033~MeV and a very small branching ratio of about 2.78 $\times$ 10$^{-6}$ \cite{Anthony02,Raabe09,Pfutzner15}. Therefore, neutron emission following the decay of $^6$He is possible only through a dark decay
\begin{equation}
\mathrm{^{6}He} \rightarrow \mathrm{^{4}He} + n + \chi,
\end{equation}

which translates into
\begin{equation}
    937.993 < m_{\chi} < m_n - 0.975 \; \mathrm{(MeV)}.
\end{equation}
As such, this decay probes a significant ($\sim$38\%) part of the total $m_{\chi}$ range, and the vast majority ($\sim$76\%) when $\chi$ is a dark matter candidate ($m_{\chi} < m_p+m_e$). In order to resolve the neutron lifetime discrepancy, a simple estimate \cite{pfutzner18} arrived at an equivalent branching ratio of $1.2\times10^{-5}$ for $^6$He, which we discuss in greater detail below.

The $^6$He isotopes were produced at the Grand Acc\'el\'erateur National d’Ions Lourds (GANIL) SPIRAL1 facility \cite{Villari01}.
After mass separation by a dipole magnet, the $^6$He$^+$ beam was guided at 25~keV to the low energy beam line LIRAT and was implanted in a thin 150~$\mu$m aluminum catcher. A fast electric deflector located upstream from the last dipole magnet was used to chop the beam on and off. In a typical cycle, the ions were implanted during $t_{\textsc{ON}} = 3$~s followed by 7~s of beam off used to observe $^6$He decays. The same configuration was used for a beam of $^8$He$^+$, a ${\beta}^-$ delayed neutron emitter, serving as a reference for the detection setup characterization. During beam tuning, a movable silicon detector was inserted into the LIRAT line in front of the detection section to measure the incoming beam intensity, reduced by a reduction grid of known reduction factor.

\begin{figure}
    \centering
    \includegraphics[width=1.\columnwidth]{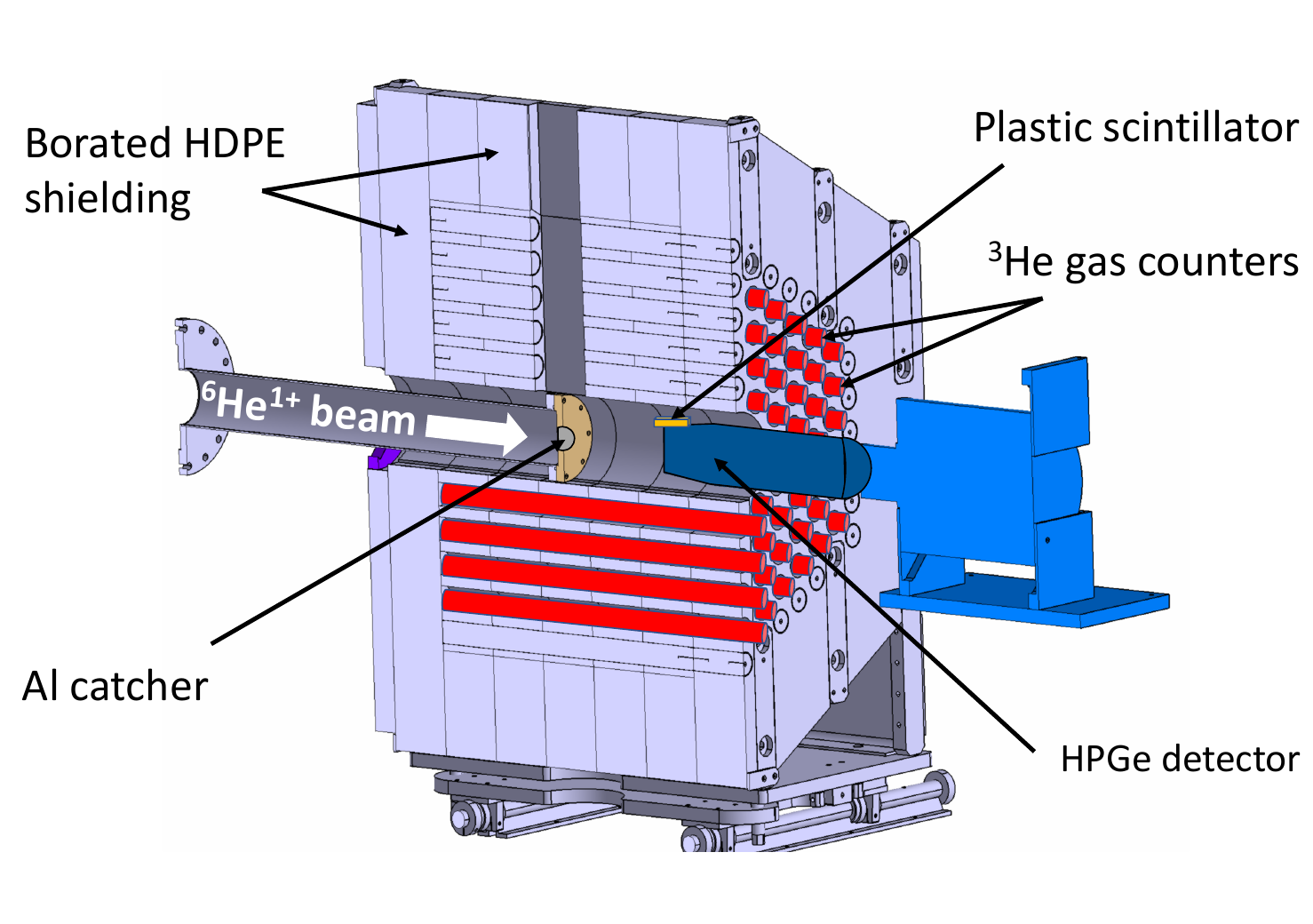}
    \caption{Sectional view of the detection array optimized for neutron detection, including the 4$\pi$ counter TETRA, a HPGe detector and a small solid angle plastic scintillator.}
    \label{fig:tetra_setup}
\end{figure}

Since the branching ratio is expected to be very low ($< 1.2\times10^{-5}$) and the neutron energy to be detected is smaller than 1~MeV, a high efficiency neutron detection setup with low energy neutron threshold is essential. 
We used the 4$\pi$ $^3$He neutron counter TETRA (see Fig.~\ref{fig:tetra_setup}) consisting of 82 counters filled with a gas of $^3$He and a 1$\%$ admixture of CO$_2$ at a nominal pressure of 7~atm~\cite{Terakopian81,Testov16}. Neutrons are detected thanks to the energy deposition of the proton and triton reaction products resulting from neutron capture by $^3$He nuclei.
The total neutron detection efficiency obtained using the Monte Carlo N Particle transport code is almost constant and close to 60\% from thermal neutron energies up to 0.6~MeV~\cite{Testov16}, which corresponds to the energy domain expected for a free neutron emission resulting from a dark decay in $^6$He.
Using a $^{252}$Cf source (average neutron energy at $\approx2.3$~MeV) located at the center of the detector, an efficiency of $\varepsilon_{n}$ = 53(5)$\%$ was measured.
This efficiency, with an uncertainty dominated here by the activity of the source, is consistent with the one reported in Ref.\cite{Testov16} and was conservatively used as a reference in the following data analysis.
Two other detectors were used during the experiment: a HPGe (crystal with a 6.3~cm diameter and a length of 5.6~cm)  located downstream the catcher to detect $\gamma$-rays and a plastic scintillator ($0.5\times0.5\times1.5$~cm$^3$) placed on the top of the HPGe to monitor $\beta$-particle emission. The HPGe detector was calibrated using a $^{152}$Eu source.

Individual signals from all detectors were sent to the input channels of the digital data acquisition FASTER \cite{Faster}. Five different types of data were recorded: i) the resulting pulse height of signals from the $^3$He counters after processing by the TETRA electronic chains (preamplifier-shaper-amplifier) \cite{Testov16}, ii) the pulse height of the HPGe signals after charge preamplification and a digital filtering by the FASTER system, iii) the charge from the plastic scintillator photomultiplier tube, iv) the pulse rate providing the intensity of the primary beam on the SPIRAL1 target and v) a pulse providing the start time of each implantation/decay cycle. All events were individually time stamped with a 8~ns resolution, allowing on-line and off-line correlations between all channels and, more importantly, to determine the time $t_c$ of each event within the beam cycle. All the on-line thresholds were set as low as possible at around 1~keV for each counter of TETRA and 10~keV for the plastic scintillator. The best individual thresholds, in particular for neutron-photon discrimination, could therefore be adjusted off-line for each detector. 
Before and after the experiment, data were recorded under the experimental conditions using $^{252}$Cf and $^{152}$Eu sources to determine the response functions of the detectors. During the experiment, the plastic scintillator was located at two different distances from the beam catcher (21~cm and 43~cm) to cope with the different detection rates for $^8$He and $^6$He runs. Finally, to check the background level, long runs were performed with the primary beam hitting the SPIRAL1 target but without sending any exotic nuclei in the LIRAT beam line.

To test and to quantify the response of the entire detection setup, we also performed the decay study of $^8$He \cite{Bjornstad81,Borge86,Borge93,Mianowski10} which has the advantage of being a ${\beta}$, neutron and $\gamma$ emitter. The half-life of $^8$He is 119(1)~ms and its decay proceeds via Gamow–Teller transitions to 1$^{+}$ excited states of $^8$Li with $Q_\beta$ = 10.664~MeV. About 84$\%$ of the  $\beta$ decays feed the first excited state in $^8$Li at the excitation energy of 980.8~keV. The remaining 16$\%$ feed broad resonances in $^8$Li followed by neutron emission (neutron energy $\sim1$~MeV) to the ground and first excited states of $^7$Li. To switch from one nucleus to the other, the magnetic field of the dipole magnet for mass separation was scaled to the appropriate mass over charge ratio.

All $^{3}$He counters from TETRA were calibrated in energy with a linear model using the full energy peak at 765~keV from the neutron capture reaction. Counters showing an anomalous response during background runs and high energy noise (appearing above 900~keV) correlated with the beam implantation cycle were discarded. Finally, 72 out of 82 counters were used. To suppress the contribution of after-pulses and of high multiplicity events due to cosmic background, multiple TETRA triggers occurring within less than 300~$\mu$s were also removed from the data analysis.
The mean $^8$He implanted rates were computed independently using the tabulated branching ratios of the $\beta n$ decay branch with TETRA and the $\gamma$ line at 980.8~keV from the de-excitation of $^8$Li with the HPGe detector. These values differ by less than $10\%$ and are in agreement with the uncertainties on both detection efficiencies and branching ratios. 

For all detectors, histograms of rates as a function of the time $t_c$ within the cycle were produced. By fitting these histograms for the neutrons detected with TETRA and for the $\beta$ particles from $^8$He and its $^8$Li filiation detected with the plastic scintillator, we checked that the resulting half-life values were consistent with the literature. The 980.8~keV $\gamma$ line was used as a reference to compute the $\beta$-detection efficiency. The number of events detected by the plastic scintillator scaled with the HPGe provides detection efficiencies of $\varepsilon_{\beta,C} = (1.02\pm0.05)\times10^{-3}$ at the closest position from the beam catcher and $\varepsilon_{\beta,F} = (2.05\pm0.11)\times10^{-4}$ at the farthest position. These values were then corrected by a factor assessed with G4beamline simulations \cite{G4Beamline} which accounts for the difference in the detection efficiencies due to the difference of the $Q_{\beta}$ values for the $^6$He and $^8$He/$^8$Li decays. For the $^6$He $\beta$ decay, the final values are $\varepsilon_{\beta,C} = (8.67\pm0.44)\times10^{-4}$  and $\varepsilon_{\beta,F} = (1.57\pm0.09)\times10^{-4}$. 
\begin{figure}[!t]
    \includegraphics[width=1.\columnwidth]{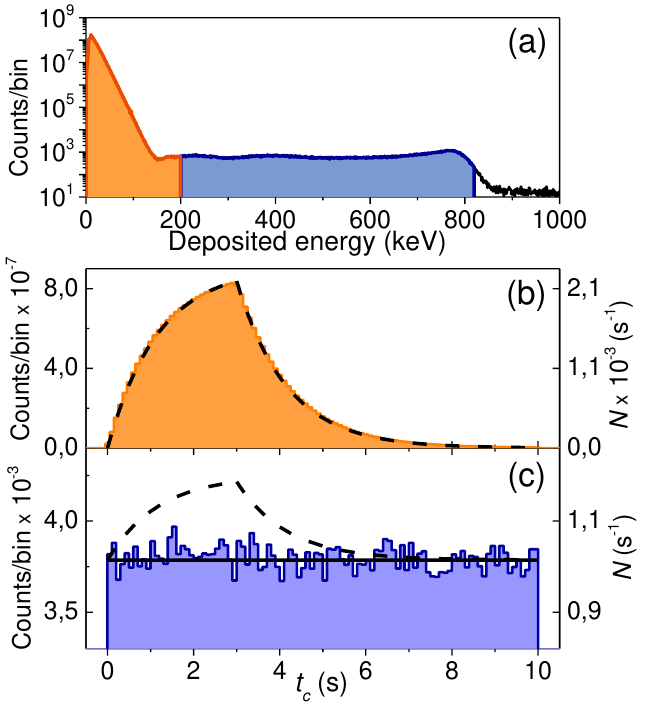}
    \caption{Deposited energy spectrum for all selected $^3$He counters (a). The orange (from 0 to 200~keV) and blue (from 200 to 820~keV) areas indicate the energy ranges where bremsstrahlung and neutron events dominate, respectively. Histograms as a function of the time in the cycle $t_c$ for events corresponding to the orange selection window~(b) and to the blue selection window~(c). The vertical scales indicate both the counts per 100~ms bin and the average detection rate $N$ as a function of $t_c$. The dashed line in panel (b) results from the fit of bremsstrahlung events with Eqs. (3) and (4) using the known lifetime of $^6$He. The plain and dashed line in panel (c) show the expected neutron detection rate assuming a dark neutron decay with a branching ratio of 0 and $2\times10^{-9}$, respectively.}
    \label{fig:tetra_response}
\end{figure}

For runs with $^6$He$^+$ ions, a total of 37374 cycles of 10 seconds duration were used for the analysis. The total number of implanted $^6$He during the experiment, deduced from the plastic scintillator data is $(1.21\pm0.06)\times10^{13}$ with an average implanted rate of $r_0 = (1.08\pm0.06)\times10^{8}$~pps within the three implantation seconds per cycle. Assuming no neutron dark decay, the neutron detection with TETRA should only arise from ambient background at a constant rate. Otherwise, for a hypothetical neutron dark decay in $^6$He, an excess of free neutrons emitted at a rate proportional to the implanted and surviving $^6$He population would be detected. Such a signal can be unambiguously evidenced by fitting the neutron detection rate $N(t_c)$ averaged over all cycles as a function of $t_c$, with the following functions :
\begin{equation}
    N_1(t_c) = \phi (1-e^{-t_c/\tau}) + b \mathrm{,}
\end{equation}
\begin{equation}
   N_2(t_c) = \phi (e^{t_{\textsc{on}}/\tau}-1) e^{-t_c/\tau} + b \mathrm{,}
\end{equation}
where $N_1(t_c)$ is the detection rate during the implantation period for $0 \leq t_c \leq t_{\textsc{ON}}$ and $N_2(t_c)$ the detection rate in the beam off period for $t_{\textsc{ON}} \leq t_c \leq 10 s$. The \textit{b} parameter corresponds to the constant background rate, $\tau$ is the $^6$He lifetime fixed at 1.164~s \cite{Kanafani22} and $\phi$ corresponds to the average $^{6}$He implantation rate $r_0$ during the beam on period, scaled by the neutron detection efficiency and the dark decay branching ratio Br$_{\chi}$.

The deposited energy spectrum of all events detected by the TETRA detector for $^6$He runs is shown in Fig.\ref{fig:tetra_response}(a) with a 1~keV bin width. Neutron events correspond to a deposited energy between about 150 and 820~keV. The large number of events below 150~keV is due to the detection of bremsstrahlung radiation from $\beta$ particles emitted by $^6$He nuclei and interacting with the surrounding matter. The rate histogram versus $t_c$ for a selection of such events, with deposited energy below 200~keV (orange area of Fig.\ref{fig:tetra_response}(a)), is shown in Fig.\ref{fig:tetra_response}(b) with a 100~ms bin width. The detection rate follows, as expected, the decay rate of the $^6$He nuclei implanted in the catcher. For events corresponding to the region colored in blue in Fig.\ref{fig:tetra_response}(a), where neutron signals dominate, the rate histogram versus $t_c$ is shown in Fig.\ref{fig:tetra_response}(c). This histogram does not display visible time dependence and can therefore be mostly attributed to the ambient background. To show the effect of a dark decay in $^6$He, the expected neutron detection rate assuming a branching ratio of $2\times10^{-9}$ is also indicated by a dashed line in Fig.\ref{fig:tetra_response}(c).

\begin{figure}[!hbt]
    \includegraphics[width=1.\columnwidth]{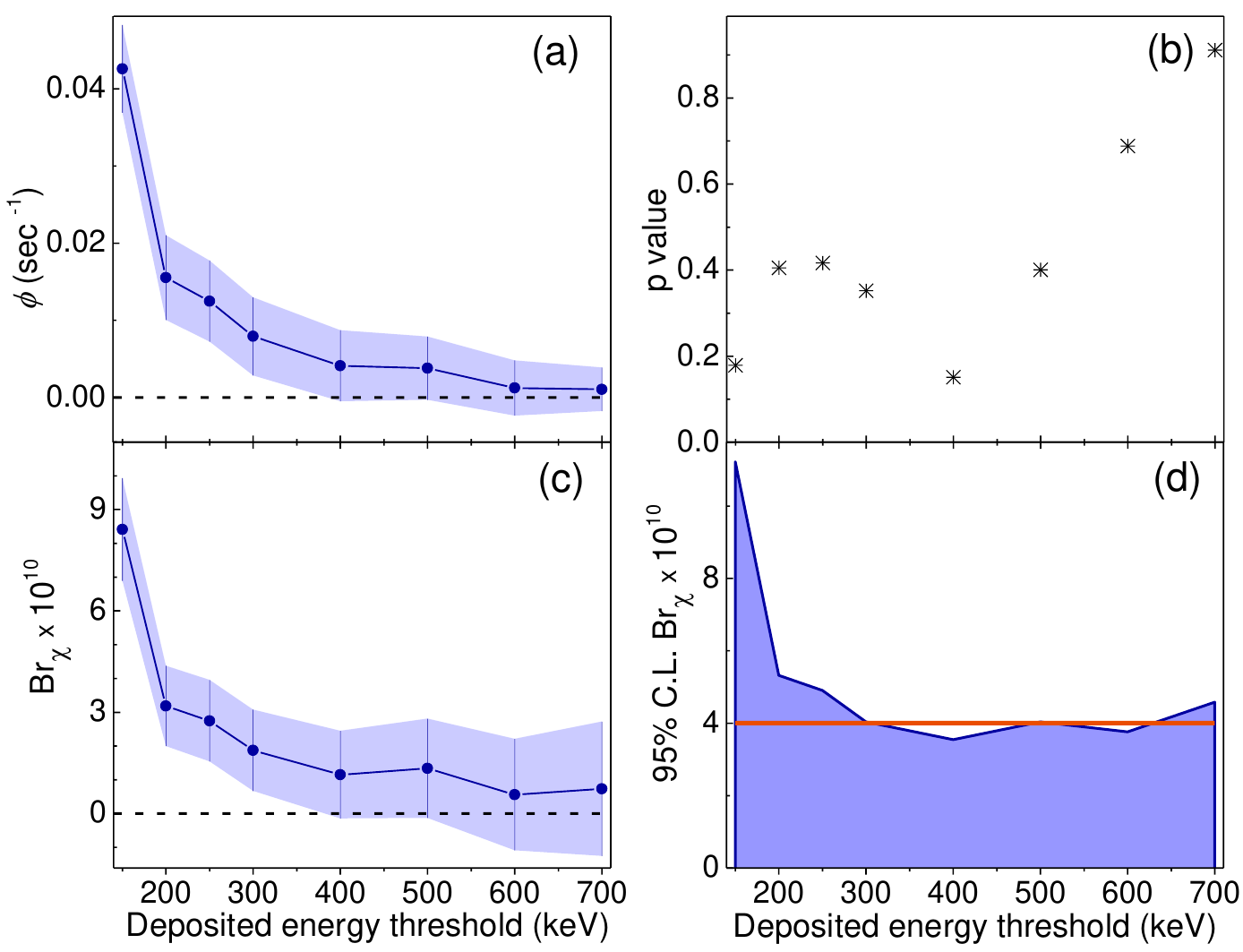}
    \caption{(a) $\phi$ values resulting from fits with Eqs. (3) and (4) as a function of the deposited energy threshold. Error bars are purely statistical. (b) Corresponding $p$-values for each fit. (c) Resulting branching ratio after scaling by the neutron detection efficiency and the mean implanted rate of $^6$He. Error bars account for both the statistical and systematic uncertainties. (d) Branching ratio limits for the dark decay at a 95\% confidence level.}
    \label{fig:best_fit}
\end{figure}

To separate a dark decay signature from a possible contamination by bremsstrahlung events within the selected detection energy range, fits with Eqs.(3) and (4) were performed on multiple data sets sharing a common upper limit at 820~keV but with lower limits ranging from 150~keV to 700~keV. The neutron detection efficiency $\varepsilon_n$ of TETRA in these deposited energy limits goes from $46.89\pm5.09\%$ down to $13.18\pm1.43\%$ using 72 counters.  The values of the $\phi$ parameter and their statistical errors resulting from the fits are displayed in Fig.\ref{fig:best_fit}(a). Note that these results are not statistically independent as the different data selections share the same upper limit in deposited energy. The corresponding $p$-values (goodness-of-fit) in Fig.\ref{fig:best_fit}(b) are all well above 0.05 showing a good statistical consistency between the data and the fit model. The corresponding branching ratios for a neutron dark decay, Br$_{\chi}$, are shown in Fig.\ref{fig:best_fit}(c). They were inferred from the $\phi$ values, scaled by the neutron detection efficiency for the corresponding deposited energy range and the $^6$He implantation rate. The error bars account for both statistical and systematic uncertainties (on $\varepsilon_{\beta}$ for the plastic scintillator and on $\varepsilon_{n}$ for each deposited energy threshold in TETRA) that were summed quadratically. The first point at 150~keV exhibits a clear sign of a bremsstrahlung contribution that quickly decreases and stabilizes at higher energy thresholds. For energy thresholds larger than 300~keV, Br$_{\chi}$ values are compatible with zero. The error bars for each point are largely dominated by the statistical uncertainty. To further check the robustness of the analysis, the fits of the rate histograms versus $t_c$ were also performed with different bin widths ranging from 10~ms to 500~ms. The suppression of multiple TETRA triggers within less than 300~$\mu$s was also cancelled or extended to 1~ms. These tests did not cause any significant change in the fit results. Finally, the 95\% confidence level upper limits of Br$_{\chi}$ as a function of the detection energy threshold are shown in Fig.\ref{fig:best_fit}(d). 
Considering energy thresholds above 300~keV, where bremsstrahlung contamination vanishes, the following conservative 95\% C.L. upper limit was obtained :
\begin{equation}
    \mathrm{Br}_{\chi} \leq 4.0\times10^{-10}.
    \label{eq:BR_result}
\end{equation}

\begin{figure}
    \centering
    \includegraphics[width=1.\columnwidth]{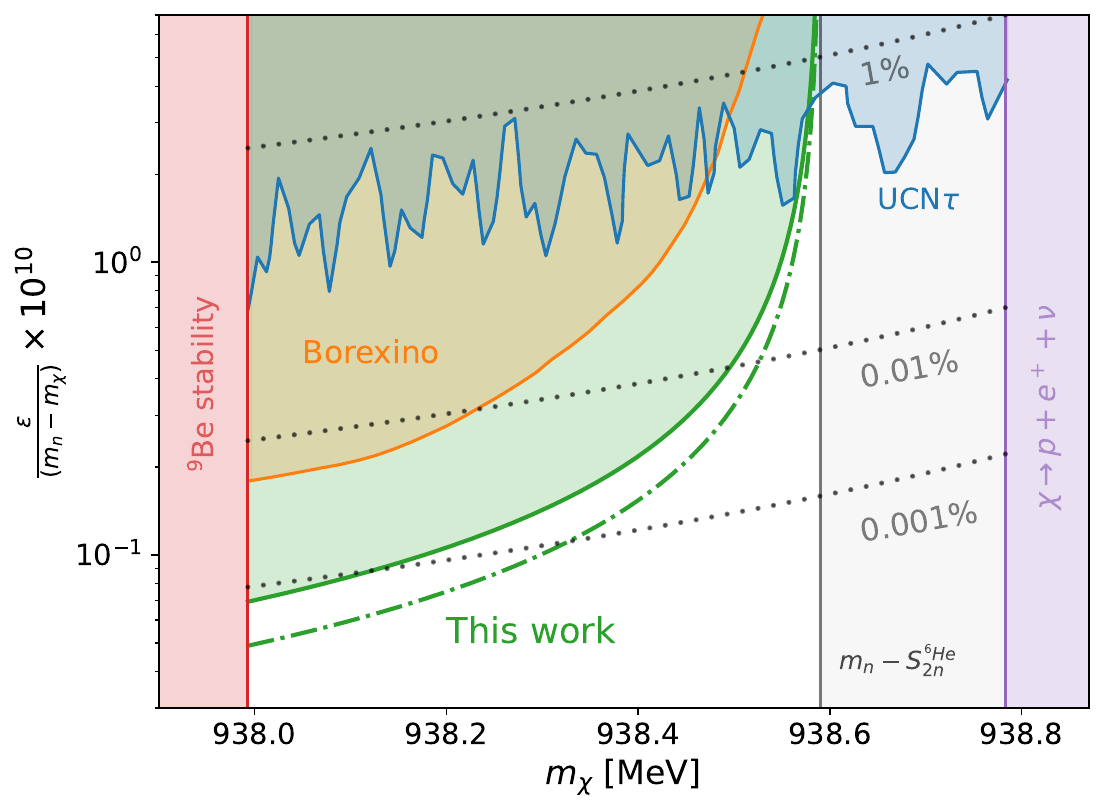}
    \caption{Comparison of competitive constraints (UCN$\tau$ \cite{tang2018}, Borexino \cite{Agostini2013, McKeen2023}) with the results of this work for Fornal and Grinstein's first model \cite{fornal18}. Dotted lines refer to fractions of the free neutron decay rate into a dark decay, $n\to \chi + \gamma$, with $1\%$ the difference between beam and bottle life times. We assume a conservative estimate (solid) of the nuclear matrix element for $^6$He dark decay and complete overlap (dash-dot) of the wave functions. In both cases, we improve global constraints by one to several orders of magnitude depending on $m_{\chi}$.}
    \label{fig:exclusion1}
\end{figure}

We may connect this result directly to the models proposed in the original work by Fornal and Grinstein \cite{fornal18}, where $n\to\chi + \gamma$, for the decay of $^6$He $\to$ $^4$He $+$ $n+\chi$. In its most basic iteration, the Lagrangian contains just two parameters - a mass mixing $\varepsilon$ and dark fermion mass $m_\chi$ - and the dark decay depends on their dimensionless combination $\varepsilon/(m_n-m_\chi)$. We note that, contrary to Ref. \cite{pfutzner18}, the $^6$He dark decay requires a calculation of the three-body phase space after which we may transform the upper limit of Eq. (\ref{eq:BR_result}) into the result in Fig. \ref{fig:exclusion1}. In doing so we consider the overlap between initial and final state wave functions, probing the degree to which one of the halo neutrons can be considered free, from unity to just 0.5. The latter corresponds to a conservative estimate, with experimental results on the order of 0.75-0.85 \cite{Ejiri2019, Winfield2001, Goss1975}. In both cases, our results show an improvement by one to several orders of magnitude compared to the state of the art, depending on $m_{\chi}$. Indeed, this work puts constraints on branching ratios of $\mathcal{O}(10^{-5})$ of the free neutron for select $m_{\chi}$, compared to $1\%$ required to resolve the beam-bottle discrepancy.

The increase in precision reported here means that, unlike searches for discrete photons \cite{tang2018} or pair production \cite{sun2018, Klopf2019}, our results can put constraints on more elaborate models with multiple dark partners such as those tested in neutron stars. The clean decay signature of a halo nucleus highlights the potential for this method, and paves the way for further studies with additional isotopes.

\begin{acknowledgments}
We gratefully acknowledge the support of the  accelerator staff and the technical staff of GANIL. We thank the JINR Dubna for the use of the TETRA detector 
This work was supported by the collaboration agreements IN2P3 - JINR Dubna Numbers 21-98 and 21-5003TMP.
  
\end{acknowledgments}

\providecommand{\noopsort}[1]{}\providecommand{\singleletter}[1]{#1}%

\end{document}